\documentclass[paper]{JHEP}
\usepackage{epsfig}
\newcommand{\MYEPSFIGURE}[3][v]{\begin{figure}[#1]\begin{center}\epsfig{file=#2}\end{center}\caption{#3}\end{figure}}
\newcommand{\MYTABULAR}[4][v]{\begin{table}[#1]{\begin{center}\begin{tabular}{#2}#3\end{tabular}\end{center}}\caption{#4}\end{table}}
\def\ifmath#1{\relax\ifmmode#1\else$#1$\fi}
\def\etal{{\it et al.,}}

\def\ie{{\it i.e.}}

\def\cf{{\it cf.}}
\def\ra{\rightarrow}
\def\bc{\begin{center}}
\def\ec{\end{center}}
\def\bi{\begin{itemize}}
\def\ei{\end{itemize}}
\def\bv{\begin{verbatim}}
\def\ev{\end{verbatim}}
\newcommand{\beq}{\begin{equation}}
\newcommand{\eeq}{\end{equation}} 
\newcommand{\beqa}{\begin{eqnarray}}
\newcommand{\eeqa}{\end{eqnarray}}
\def\CP     {\ifmath{C\!P}}
\def\babar{\mbox{\sl B\hspace{-0.4em} {\scriptsize\sl A}\hspace{-0.4em} B\hspace{-0.4em} {\scriptsize\sl A\hspace{-0.1em}R}}}
\def\KS    {\ifmath{K^0_{\scriptscriptstyle S}}} %<<<
\def\Kbar{\ifmath{\kern 0.2em\overline{\kern -0.2em K}}{}}   %<<<new
\def\Bbar{\ifmath{\kern 0.18em\overline{\kern -0.18em B}}{}} %<<<new
\def\Dbar{\ifmath{\kern 0.2em\overline{\kern -0.2em D}}{}}   %<<<new
\def\jpsi  {\ifmath{{J\mskip -3mu/\mskip -2mu\psi\mskip 2mu}}}%<<<
\def\Y#1S{\ifmath{\Upsilon\rm(#1S)}} %<<<new

\def\kev  {\ifmath{\mbox{\,ke\kern -0.08em V}}} %<<<
\def\mev  {\ifmath{\mbox{\,Me\kern -0.08em V}}} %<<<
\def\gev  {\ifmath{\mbox{\,Ge\kern -0.08em V}}} %<<<
\def\gevc {\ifmath{\mbox{\,Ge\kern -0.08em V$\!/c$}}} %<<<
\def\mevc {\ifmath{\mbox{\,Me\kern -0.08em V$\!/c$}}} %<<<
\def\gevcc{\ifmath{\mbox{\,Ge\kern -0.08em V$\!/c^2$}}} %<<<
\def\mevcc{\ifmath{\mbox{\,Me\kern -0.08em V$\!/c^2$}}} %<<<

\def\ps   {\ifmath{\,\mbox{ps}}}

\def\ssa  {\ifmath{\sin 2 \alpha }}
\def\ssb  {\ifmath{\sin 2 \beta }}
\newcommand {\deltamd}{\ifmath{{\Delta {m}_{B_d}}}}
\newcommand {\deltams}{\ifmath{\Delta {m}_{B_s}}}

\newcommand {\vbu}{\ifmath {|V_{ub}|}}
\newcommand {\vbc}{\ifmath {|V_{cb}|}}
\newcommand {\vbuvbc}{\ifmath |\frac{\vbu}{\vbc}|}
\newcommand {\mvbu}{\ifmath <\vbu>}
\newcommand {\epsk}{\ifmath |\epsilon_K|}
\newcommand {\fbb}{\ifmath { f_{B_d} \sqrt{B_{B_d}} }}
\newcommand {\fbbs}{\ifmath {f_{B_s} \sqrt{B_{B_s}}}}
\newcommand {\bk}{\ifmath {B_K}}
\newcommand {\param}{\ifmath{A,\bar \rho,\bar \eta}}

\newcommand {\rhobar}{\ifmath {\bar{\rho}}}
\newcommand {\etabar}{\ifmath {\bar{\eta}}}

\newcommand {\sss}{\ifmath {(\ssa,~\ssb)}}
\newcommand {\xis}{\ifmath {\xi_s}}
\newcommand {\epsK} {\ifmath {|\varepsilon_K|}}
\newcommand {\aPK} {\ifmath {a_{\psi K_S}}}
\newcommand {\epseps} {\ifmath{ \varepsilon^\prime/\varepsilon}}
\newcommand {\pdf} {{\em p.d.f.}}
\def \bsbsbar{{\hbox{$B_s$}}{\hbox{\kern -.45cm\lower -.3cm
\hbox{${\scriptscriptstyle(}{\scriptstyle -}{\scriptscriptstyle
)}$}}}}

\title{Overall determination of the CKM matrix}

\author{St\'ephane Plaszczynski\thanks{Talk given at Heavy Flavours 8, Southampton, UK, 1999}~
and Marie-H\'el\`ene Schune\\ Laboratoire de l'Acc\'el\'erateur
Lin\'eaire,IN2P3-CNRS et Universit\'e de Paris-Sud, F-91405 Orsay \\
E-mails: {\tt plaszczy@lal.in2p3.fr, schunem@lal.in2p3.fr} }

\abstract{ We discuss the problem of theoretical uncertainties in the 
combination of observables related to the CKM matrix elements and
propose a statistically sensible method for combining them. 
The overall fit is performed
on present data, and constraints on the matrix elements are 
presented as well as on \fbb. We then explore the
implications of recent measurements and developments: \jpsi\KS\ CP
asymmetry, \epseps\ and $B \ra K \pi$  branching fractions. Finally, we
extract from the overall fit the Standard Model expectations for the
rare kaon decays $K \ra \pi \nu \bar \nu$.
}

\preprint{LAL 99-67}

\begin{document}

\newpage
\section{Introduction}
The Cabbibo-Kobayashi-Maskawa (CKM) matrix is extensively studied nowadays.
 With the birth of the new B factories
and the upgrade of the Tevatron experiments, high precision measurements in the 
B meson sector are expected and 
the question of testing the CKM ansatz is pushed towards more and more 
stringent limits.
By ``testing'' we mean two aspects:
\begin{enumerate}
\item Given that the Standard Model (SM) is right, what is the best
knowledge we have on the CKM free parameters?
\item Are all the measurements involving CKM matrix elements compatible
within their errors?
\end{enumerate}
As we shall see (section \ref{sec:pb}), a key point in performing this test 
is a proper treatment of the theoretical estimates that enter the
description of the observables.

First attempts to combine several observables were performed by simply
drawing in the Unitarity Triangle (UT) plane individual 95\% CL regions 
for $\rho$ and $\eta$,
obtained by varying coherently the experimental and theoretical
errors of each observable. These regions had the advantage of being
geometrically simple\footnote{Note that this is no longer the case
when considering for instance $\ssa$.}. 
The intersection of these regions was taken as a 95\% CL for ($\rho,\eta)$.
While statistically wrong (it neglects the correlations induced by the 
combination), this method gives surprisingly good results. It can
certainly be used to get an insight into the effect on CKM
parameters of a given set of observables.

More sophisticated fits have been proposed \cite{ali, roudeau}
but, as in the previous case, assuming some flat (or Gaussian)
distribution for theoretical parameters.

We propose here a way of decoupling the experimental measurements from
the theoretical estimates (section \ref{sec:sol}). It will be
illustrated in the (\rhobar,\etabar) plane  
and the $\sss$ plane (section \ref{sec:1999}). 
The overall fit leads also to
constraints on the theoretical parameters \fbb\ and \bk~
(section \ref{sec:fbb}).
Recent experimental and theoretical developments are then
investigated (section \ref{sec:recent}). Finally, the impact of B
factories is discussed  and the SM rare $K$ branching ratios 
are extracted (section \ref{sec:future}).

\section{Outline of the method}

\subsection{Least Squares method}

The method exposed here is discussed in more detail in \cite{bbook}. It can
naturally accommodate any new measurements (as in the case of the
Fleischer-Mannel bound \cite{rfm}).\\

The CKM matrix contains 4 independent parameters, which can be taken as
three angles and a phase, or, exhibiting a hierarchy,
as the (improved) Wolfenstein parameters: $\lambda,\param$.
The Cabbibo angle is already very well known \cite{PDG} and $\lambda= \sin
\theta_c=0.2205$ will be considered
as fixed in the following.

Our task is therefore to determine 3 independent parameters, given a
set of measurements $Y_i = <Y_i> \pm \sigma_i$ and their theoretical
description Y(\param).

Considering the two aspects discussed in the introduction, 
we are naturally lead to use the least squares estimate method. One builds:
\beqa
\label{eq:chi2}
\chi^2(\param) = \sum_i \left[ \frac{<Y_i>-Y(\param)}{\sigma_i}
\right]^2
\eeqa
Minimising the $\chi^2$ function:
\begin{enumerate}
\item the values of (\param) taken at the minimum of the function
provide the least square estimates. Hyper-regions (as the ($\rhobar,\etabar$)
projection, \ie~the UT) at a given confidence level can be constructed.
\item To test the compatibility between all measurements (and their
theoretical description), one studies the value taken by the function
at its minimum, $\chi^2_{min}$. When the errors are Gaussian, one can
further use the $\chi^2$ probability distribution to quantify it.
\end{enumerate}

\subsection{\label{sec:pb}The problem with theoretical estimates}

Life would be simple if 
factors that include some level of model dependence
did not enter into the calculation. In the 
following, these will the bag factors for $K$ and $B$ mesons ($\bk,
B_B$), the $B$ decay constant ($f_B$, which will be combined with the
previous bag factor to give \fbb) and a large part of the error related to 
$\vbu$.

While the error on a measured quantity has a clear statistical
meaning\footnote{To some level one may argue that systematic errors have
also unknown \pdf\ - whenever possible we will therefore include this
part of the error in the ``model-dependent''term.}, this is not the case
for the errors quoted by theorists for their
estimates. Here enters a level of subjectivity reflecting a
``degree of belief'' that the true value lies inside some range
\cite{stone}. 
This however may vary from one person to another and is impossible to
quantify in terms of a probability distribution function (\pdf).

Furthermore, not knowing what the \pdf\ of a given theoretical estimate
is makes it impossible to combine with the other estimates 
on the market.

Finally, a crucial point is that, {\em not knowing where the true
value of a parameter lies is not equivalent to taking a flat 
probability distribution within some bounds}. This is not a matter of 
philosophical discussion; any \pdf\ extracted from a combination using a 
flat \pdf\ for theoretical estimates is senseless.

\subsection{\label{sec:sol}One solution}
One method to overcome these problems is the following:
\bi
\item Determine a ``reasonable range'' for each theoretical parameter, 
given the spread of published results. Here reasonable might mean
``conservative''.
\item Bin the whole range, and scan all the values of the theoretical
parameters. For each set of values (a ``model''), build and minimise 
the $\chi^2$
(equation~\ref{eq:chi2}) using statistical errors only. Since these are
Gaussian, one can test the compatibility between the measurements {\em 
for this set of theoretical values}. The estimate is rejected if it
does not pass a $Prob(\chi^2)$ test. If it succeeds, draw a 95\% CL contour in 
the UT plane.
\item Now, one does not know which contour is right, but presumably
one of them will be correct
(assuming the scanned range is really reasonable). Therefore our
maximum knowledge is that the set of all these contours is an overall
95\% CL. Visually, it corresponds to the {\em envelope of all the
individual contours}.
\ei
The only output of such a method is to provide an overall 95\% CL
region in the ($\rhobar,\etabar$) plane 
(or in \sss). No particular \pdf\ can be inferred, 
and estimates of the mean and the ``$\sigma$'' have no particular meaning;
the \pdf\ within the envelope is simply unknown.

\section{\label{sec:1999} CKM 1999}
\subsection{Observables}

Our present knowledge of the CKM elements lies in the following observables.

\subsubsection{\vbc}

\vbc\ gives a direct access to the $A$ parameter of the Wolfenstein
parametrization. Much work on the experimental/theoretical side
allows us to quote \cite{calvi}:
\beqa
\label{eq:vbcval}
\vbc = 0.040 \pm 0.002 
\eeqa
Note that part of the error quoted here is indeed model-dependent,
but it was checked that no ``visual'' difference can be observed when
treating it as a model-dependent range. We can safely consider it as
statistical in the following. 

\subsubsection{\vbu}

\bi
\item there has been a recent update of the CLEO $B\ra \rho \ell \nu$
analysis \cite{rhoelnu} giving:
\beqa
\vbu = [3.25 \pm 0.14(stat) ^{+0.21}_{-0.29}(syst) 
\nonumber \\ \pm 0.55 (model)] \times 10^{-3}  \nonumber
\eeqa
\item also LEP reports a combined value \cite{calvi}:
\beqa
\vbu= [4.05 ^{+0.39}_{-0.46} (stat+det) ^{+0.43}_{-0.51}(b \ra c) 
\nonumber \\
^{+0.23}_{-0.27}(b \ra u)\pm 0.16 (HQE)] \times 10^{-3} \nonumber
\eeqa
where the last three terms corresponds to systematics related to the
modelling of $b\ra c$, $b \ra u$ and to the Heavy Quark expansion,
used to relate the measured $BR(b \ra X_u \ell \nu)$ to \vbu.
\ei

The LEP
measurements have a statistical error about twice as large as CLEO,
and both experiments have the same order of magnitude for detector
systematics. But clearly the dominant part of the error is model-dependent.
Since it would be questionable to go beyond 15\% (relative) error for
this term \cite{stone} we will use as a reasonable guess
estimate\footnote{We do not use in this part the CLEO inclusive analysis since it gives
comparable results and is older.}:
$$\vbu= \mvbu \pm 0.13 \times 10^{-3}(stat) $$
where we will vary the mean within the range:
\beqa
\mvbu \in [2.9,3.9]\times 10^{-3}
\eeqa
Note that the final results do not depend crucially on the details of
the value used here.

\subsubsection{\epsK}

We use for this measurement the PDG value~\cite{PDG}:
\beqa
\epsk = (2.285 \pm 0.018) \times 10^{-3} 
\eeqa
The theoretical description of this observable in terms of CKM matrix
elements can be found in the literature (see for instance \cite{buras}) and
requires a value for the non-perturbative QCD bag factor, \bk.
Following Buras \cite{buras}, we will use: 
\beqa
\bk \in [0.65,0.95] 
\eeqa

\subsubsection{\deltamd}

The LEP oscillation working group has produced an optimal combined
value for the B mixing frequency \cite{LEPOSC}:
\beqa
\deltamd= 0.473 \pm 0.016 \ps^{-1}
\eeqa
Again, theoretical computations require not only a value for
the corresponding bag factor $B_B$, but also in this case the
$B$ decay constant $f_B$, so that
finally the relevant model-dependent theoretical parameter is \fbb.

Here again, opinions may vary on a conservative range. Following the
work performed in the \babar\ Physics book \cite{bbook}, we will use:
\beqa
\fbb \in [160,240] \mev
\eeqa
Given the large variations observed with the first unquenched estimates 
\cite{hashimoto}, this range may even be over-optimistic.

\subsubsection{\label{sec:dms} \deltams}
The study of the mixing frequency in the strange B meson sector
has not lead to a
firm measurement, but much information can be inferred from the
combination of the {\em amplitudes} performed by the LEP Oscillation
working group \cite{LEPOSC}.

As already detailed in \cite{bbook} and \cite{rfm}, we use the available information  optimally by building a $\chi^2$ of the form:
\beqa
\chi^2(\param) = \left[ \frac{{\cal A}(\deltams(\param))-1}{\sigma_{\cal A}} 
\right]^2 
\eeqa
where $ {\cal A}, \sigma_{\cal A} $ are extracted from the amplitude
curve, and \deltams(\param) is the theoretical computation.

It is frequently argued that theoretical errors cancel when
taking the ratio $\deltams/\deltamd$. This is only true to
the level of precision determined by the new theoretical parameter
\beqa
\xis^2=\left[\frac{\fbbs}{\fbb}\right]^2
\eeqa
For this parameter, we will scan the range\footnote{
Note that since we basically have a lower bound on \deltams, 
the only relevant value for the CKM combination is the upper value
$\xi_{s MAX}^2$. This factor may vary depending on the authors and, starting 
from similar ``guesstimates'', the square factor enhances the differences.}
(\cite{buras} and references therein):
\beqa
\xis^2 \in [1.12,1.48]
\eeqa

\MYTABULAR[h]{|c|c|c|}{
\hline 
Measurement & Mean value & Error \\
\hline
\vbc &  $.040$ & $0.002$ \\
$\vbu (\times ~10^{3})$ &  $\mvbu$ &$0.13$ \\
$\epsk (\times ~10^{3})$ & $2.285$ & $0.018 $ \\
\deltamd $(\ps^{-1})$& $0.473$ &$0.016 $  \\
\deltams \cf(\ref{sec:dms})& ${\cal A}$ & $\sigma_{\cal A}$\\
\hline
}{\label{tab:meas}The set of measured values used in the global fit. Whenever possible 
the error is statistical only.}

\MYTABULAR[h]{|c|c|c|}{
\hline
Parameter & Min. & Max. \\ 
\hline
$\mvbu(\times 10^3)$ & 2.9 & 3.9 \\
\bk & 0.65 & 0.95 \\
\fbb (\mev)& 160 & 240 \\
$\xis^2$ & 1.12 & 1.48 \\
\hline
}{\label{tab:range}Range scanned in the global fit of the
model-dependent theoretical parameters.}
\subsection{ Unitarity Triangle}

We build the $\chi^2(\param)$ defined in section~\ref{eq:chi2} with
all the observables 
described in the previous part, only using statistical errors (table \ref{tab:meas}). 
Each model-dependent parameter is
scanned independently within the range of table \ref{tab:range}. For each set of these values, the
$\chi^2$ is minimised using the package MINUIT \cite{MINUIT} and the estimate is
kept if it satisfies a $\chi^2$ probability cut, $P(\chi^2_{min}) \ge
0.05$. We then draw the associated 95\% CL contour in the 
$(\rhobar,\etabar)$ plane, 
for this model. Each model surviving the cut is superimposed. The
envelope of all the contours is the overall 95\% CL region
for the CKM parameters ($\rhobar,\etabar$).

From figure \ref{fig:ut99} one can extract (roughly) the projections:
\beqa
0  \le \rhobar \le 0.3 \\
0.2 \le \etabar \le 0.45
\eeqa
From such a global fit, an estimate of the third CKM parameter $A$ 
is possible. However, given the model dependency induced on that
parameter by the various observables, 
it is certainly wiser to extract 
$A$ directly from the \vbc\ measurement alone:
\beqa
A=\frac{\vbc}{\lambda^2}=0.82 \pm 0.04
\eeqa

\subsection{\label {sec:sss} (\ssa, \ssb)}

The same $\chi^2$ can be built in another basis, namely 
$(A,\ssa,\ssb)$. The same procedure is applied\footnote{For
completeness, there appears a four-fold ambiguity which is
solved by taking  the minimum value of the $\chi^2$ under 
the four hypotheses for each point \cite{bbook}.} and figure \ref{fig:ss99} 
shows the 95\% CL region in the $(\ssa,\ssb)$ plane.
From that figure we get the projections (95\% CL):
\beqa
.50 \le \ssb \le .85 \\
-.95 \le \ssa \le .50 
\eeqa
The 95\% CL regions we obtain are larger than those
reported in \cite{roudeau}, especially for the \ssa\ parameter.
This comes from a (somewhat) different choice of the parameters 
(mainly $\xi_{s MAX}^2$) but especially from a
different  treatment of the
theoretical errors; in \cite{roudeau} the 95\% CL region for \ssa~is extracted from the
``\pdf'' inferred from the fit. As detailed in section \ref{sec:pb},
we disagree with that approach.

\subsection {\label{sec:fbb} Constraints on \fbb} 

So far, we have just explored the first aspect of testing (getting
the best knowledge on the CKM parameters assuming the SM is right) and 
we turn now to the second; are the results consistent?

Here, recall the procedure; model dependent terms are scanned within a 
range and for each set of them, the $\chi^2$ is computed. The value of
the $\chi^2$ at its minimum indicates the consistency of all
measurements for that set of theoretical parameters. One can therefore
reject sets of values which are inconsistent with all the measurements
(if they were all rejected, we would conclude there is a
consistency problem, implying new physics).

Figure \ref{fig:fbb} shows a projection 
of all the scanned points
that survived the $Prob(\chi^2_{min})$ cut of 5\% during the combination
in the ($\fbb,\bk$) plane. (Note these parameters are
somewhat related by lattice computations.)

Figure \ref{fig:fbb} indicates that the Standard Model combination implies:
\bi
\item $\fbb \ge 195 \mev$
\item low values of \bk\ with large values of \fbb\ are disfavoured
\ei

\section{\label{sec:recent}Recent developments}
\subsection{\aPK~(CDF)}

CDF has reported a first measurement of the \jpsi\KS\ CP asymmetry,
which leads to the (model-independent) measurement \cite{CDF}:
\beqa
\ssb=0.79^{+0.41}_{-0.44}
\eeqa
With respect to what is already known on \ssb\ from the combination above
(figure~\ref{fig:ss99}),
it clearly does not constrain the CKM matrix elements any further.

This is however the first measurement directly related to the phases
of the  CKM matrix only and the fact that $\ssb > 0$ (95\% CL) is
strong support for the validity of the CKM description.

\subsection{\epseps\ (KTeV, NA48)}

A large value for this observable has been measured by
the KTeV and NA48 collaborations. Their combination gives \cite{eps}
\beqa
\label{eq:epsval}
\epseps = (21.3\pm2.8)\times 10^{-4}
\eeqa
The CKM description of this observable involves only $\eta$:
\beqa
\label{eq:eps}
\epseps = \eta A^2 \lambda^5 F_{\epsilon^{\prime}}
\eeqa
but the function $F_{\epsilon^{\prime}}$ includes many theoretical
factors. A crude description is \cite{jamin}:
\beqa
\label{eq:feps}
F_{\epsilon^{\prime}}&=&13\left[\frac{110 \mev}{m_s(m_c)}\right]^2\times \frac{\Lambda_{\bar{MS}}^{(4)}}{340 \mev} \nonumber \\
&&\times [ {B_6^{(1/2)}} (1-\Omega_{\eta+\eta^{\prime}}) 
-0.4 B_8^{(3/2)}\left( \frac{m_t(m_t)}{165\gev} \right)^{2.5}]  
\eeqa
The main uncertainties in this formula are related to the QCD penguins
($B_6^{(1/2)}$), electroweak penguins ($B_8^{(3/2)}$) and the strange
quark mass ($m_s(m_c)$). 

The difficulty in predicting any value for
this parameter comes from the fact that two badly known parameters
($B_6^{(1/2)}$ and $B_8^{(3/2)}$) are 
subtracted and that the difference could be as low as 0. On the other hand,
this difference cannot be {\em too large}.
And an upper bound on $F_{\epsilon^{\prime}}$ directly translates
into a lower bound on $\eta$ (equation \ref{eq:eps}). For instance 
using \cite{jamin}: $B_6^{(1/2)}=1.0\pm0.3, B_8^{(3/2)}=0.8\pm0.2,
m_s(m_c)=130\pm25 \mev$, $\Lambda_{\bar{MS}}^{(4)}=340\pm50 \mev,
\Omega_{\eta+\eta^{\prime}}=0.25\pm0.08, m_t=165\pm 5 \gev$ and
coherently varying the errors, one obtains:
\beqa
F_{\epsilon^{\prime}} \le 13.4
\eeqa
which translates into:
\beqa
\eta \ge \frac{\epseps}{13.4 \vbc^2 \lambda}
\eeqa
and using the experimental input (sections~\ref{eq:vbcval} and
\ref{eq:epsval}), allowing a 2$\sigma$ variation, one obtains:
\beqa
\eta \ge 0.32
\eeqa
Given figure \ref{fig:ut99} this would be a strong constraint on 
the UT.

Unfortunately, the formula (equation~\ref{eq:feps}) is not accurate
enough. But the message is; since the measurements provide a large 
value of $\epseps$, a firm upper bound on the theoretical parameters can 
be enough to constrain significantly the $\eta$ parameter of the CKM matrix.

\subsection {Bounds from $B \ra K \pi$}
\subsubsection{The ``Fleischer-Mannel bound''}
Fleischer and Mannel have proposed \cite{FM} a constraint on the angle
$\gamma$ of the UT, solely from the measurements of \CP-averaged branching
ratio $B \ra K \pi$:
\beqa
\sin^2 \gamma \le R
\eeqa
where
\beqa
R=\frac{\Gamma (B_d \to \pi^\mp K^\pm)}{\Gamma (B^\pm \to \pi^\pm K^0)}
\eeqa
While there has been a lot of discussion about possible theoretical 
uncertainties that would weaken this bound \cite{neubertlong}, 
it suffers mainly from the 
present CLEO measurement which gives a value of $R$ consistent with
one  \cite{bkpi}:
\beqa
R=1.11^{+0.35}_{-0.31} 
\eeqa
It will certainly worth revisiting it when more accurate measurements 
become available.

\subsubsection{The ``Neubert-Rosner bound''}

Following that idea, Neubert and Rosner  have proposed a
bound on $\gamma$ from charged $B$ decays only, in which theoretical 
uncertainties are much more under control \cite{rosner}. It relies on two
absolute branching ratios, which are used to define:
\beq
R_{\ast}=\frac{\Gamma(B^\pm \ra \pi^\pm K^0)}{2(B^\pm \ra \pi^0 K^\pm)}
\eeq
and
\beq
\bar \epsilon_{3/2}= \sqrt{2}R_{SU(3)}\tan \theta_C \left[
\frac{\Gamma(B^\pm \ra \pi^\pm \pi^0)}{\Gamma(B^\pm \ra \pi^\pm K^0)}
\right]^{\frac{1}{2}}
\eeq
where in this second formula, $\theta_C$ is the Cabbibo angle and
$R_{SU(3)}$ is a precisely known correction~\cite{neubertrecent}.
Using these observables the bound reads~\cite{grossman}:
\beqa
\label{eq:neubert}
|X_R|=|\frac{\sqrt{R_{\ast}^{-1}}-1}{\bar \epsilon_{3/2}} |\le
|\delta_{EW} -{\cos \gamma}| 
\eeqa
where $\delta_{EW}$ is calculable  in terms of Standard Model parameters:
\beqa
\delta_{EW} =(0.64\pm0.09)\times \frac{0.085}{\vbuvbc}
\eeqa
The bound in equation~\ref{eq:neubert} is discriminant provided that 
$X_R$ is (statistically) below one. While the first CLEO results were 
promising, the latest updates \cite{bkpi}  do not confirm a value 
statistically different from one:
\beqa
X_R=0.72 \pm  0.98 (exp) \pm 0.03(th)
\eeqa
As in the Fleischer-Mannel case, one waits eagerly for more precise
measurements from CLEO-III, Belle and \babar.

\section{\label{sec:future}Future measurements}
\subsection{B factories}
With the start of $B$ factories, one can expect some new measurements
of \CP~asymmetries which are related to $\ssb$ and \ssa. The stakes are
however different. Given figure~\ref{fig:ss99}:
\bi
\item $\ssb$ is already well constrained. The goal of measuring it is
to {\em test the SM}, since in a variety of models new physics may
appear only in the CKM phases \cite{nir}. It is not expected that measuring
this angle will constrain \fbb\ much more than presently.
\item $\ssa$ is largely unknown and the goal of $B$ factories is to
{\em measure} it. The extraction of that angle from the measured 
\CP\ asymmetries is
difficult (impossible?) and will certainly require several years of
running \cite{bbook}. The most promising channel is presently $B \ra 3 \pi$ in
which all the amplitudes can be extracted from a global fit to the Dalitz 
plot.
\ei

\subsection{$K \ra \pi \nu \bar \nu$}
 
Both the charged mode and the neutral one are ``theoretically clean'' 
and measuring their rate would significantly constrain the CKM matrix \cite{buras}. 
For the time being, we can
extract the expected branching ratio from the global fit by scanning
all the points in the contours of figure~\ref{fig:ut99} and keeping
the minimum and maximum value of the corresponding computed branching
ratio. One obtains (for 95\% CL):
\beqa
BR(K_L \ra \pi^0 \nu \bar \nu) &\in &[1-5]\times 10^{-11} \\
BR(K^+ \ra \pi^+ \nu \bar \nu) &\in &[4-10]\times 10^{-11} 
\eeqa

\section{Conclusions}

We want to draw the attention of the reader to the difficulties
that arise when including some theoretical estimates into an overall
combination of observables relevant to the CKM determination. This was 
also emphasised by Stone \cite{stone} and will (and already does) 
limit our understanding of the CKM
parameters.
Given that a ``theoretical'' error has an unclear statistical meaning,
we conclude that 
extracting any \pdf\ from a combination including these parameters is
just meaningless, and that no ``central values'' and ``errors'' should be
ever quoted. 

Nevertheless, we have proposed a (conservative) method to obtain some 95\% CL
regions for all CKM parameters, by separating the statistical errors
due to measurements from the systematic and model-dependent ones. From such
a combination, we obtain the 95\% CL bounds:
\bi
\item $ 0 \le \rhobar \le 0.3,~0.2 \le \etabar \le 0.45$
\item $.50 \le \ssb \le .85,~-.95 \le \ssa \le .50 $
\item $\fbb \ge 195 \mev$
\ei

Among new developments, the large value measured for $\epseps$ could
constrain $\eta$ (by a lower bound) if theoretical uncertainties
were more under control (an upper bound would be sufficient). 

$B \ra K \pi$ absolute branching ratios could constrain significantly the angle
$\gamma$ of the UT but must be measured more precisely.

Finally, from the overall combination, one can extract the expected
branching ratios:
\beqa
BR(K_L \ra \pi^0 \nu \bar \nu) &\in &[1-5]\times 10^{-11} \nonumber \\
BR(K^+ \ra \pi^+ \nu \bar \nu) &\in &[4-10]\times 10^{-11} \nonumber
\eeqa

\acknowledgments
We thank warmly Yossef Nir and Helen Quinn for critical
comments on the Neubert bound, and Marta Calvi and Matthias Neubert
for kind details about their presentations at {\it Heavy Flavours}.

%%%%%%%%%%%FIGURES%%%%%%%%%
\newpage
\MYEPSFIGURE[htbp]{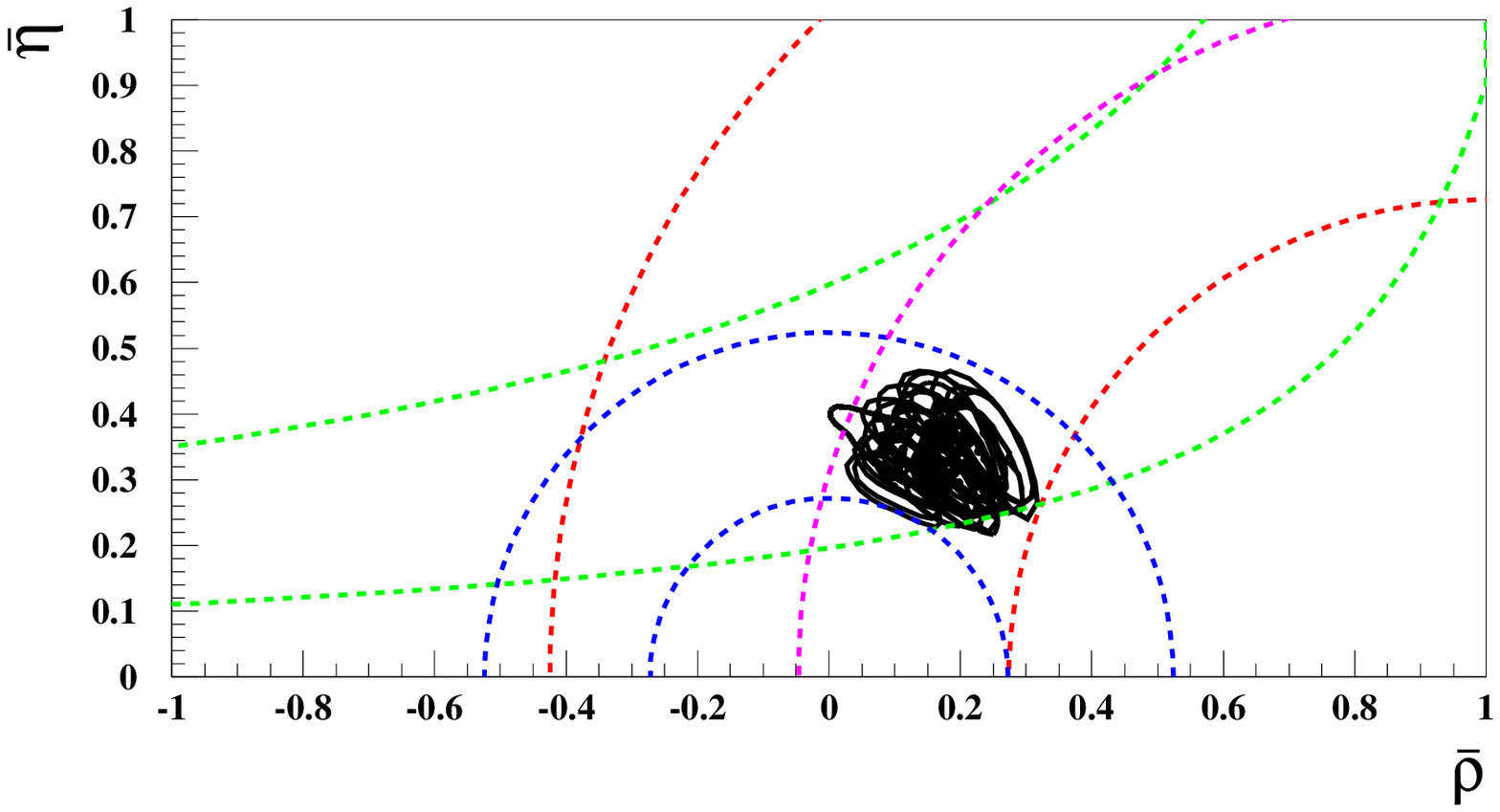,width=15cm}{\label{fig:ut99}The allowed 95\% CL region for
\rhobar-\etabar. Each contour corresponds to one theoretical model,
and the envelope of all of them is the overall 95\% CL combination.
Also shown (for historical reasons) the usual individual constraints.}
%%%%%%%%%%%%%%%%%%%%%%%%
\newpage
\MYEPSFIGURE[htbp]{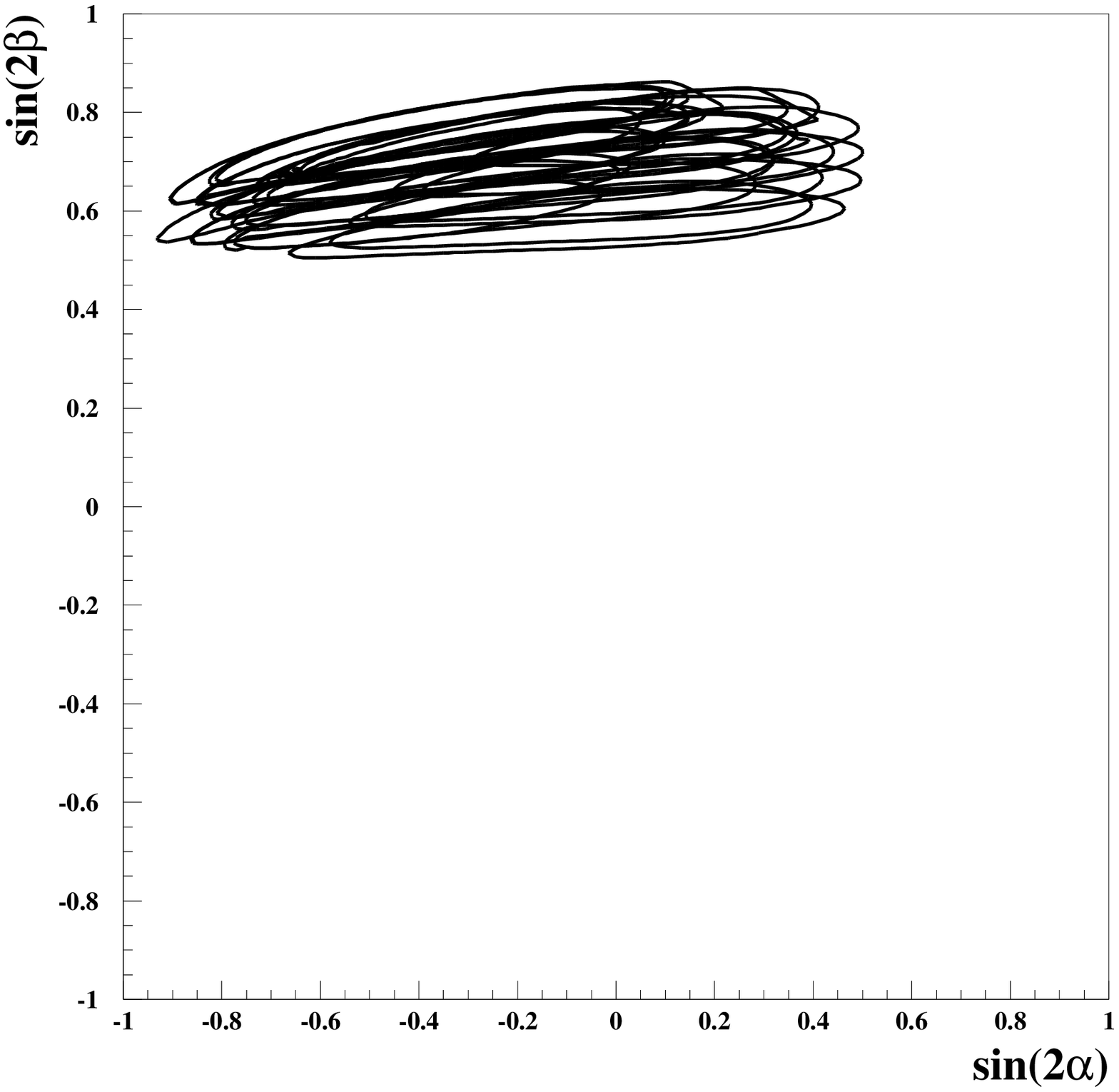,width=15cm}{\label{fig:ss99}The 95\% CL allowed region for
\ssa-\ssb. Each contour corresponds to one theoretical model,
and the envelope of all of them is the overall 95\% CL combination.}
%%%%%%%%%%%%%%%%%%%%%%%%
\newpage
\MYEPSFIGURE[htbp]{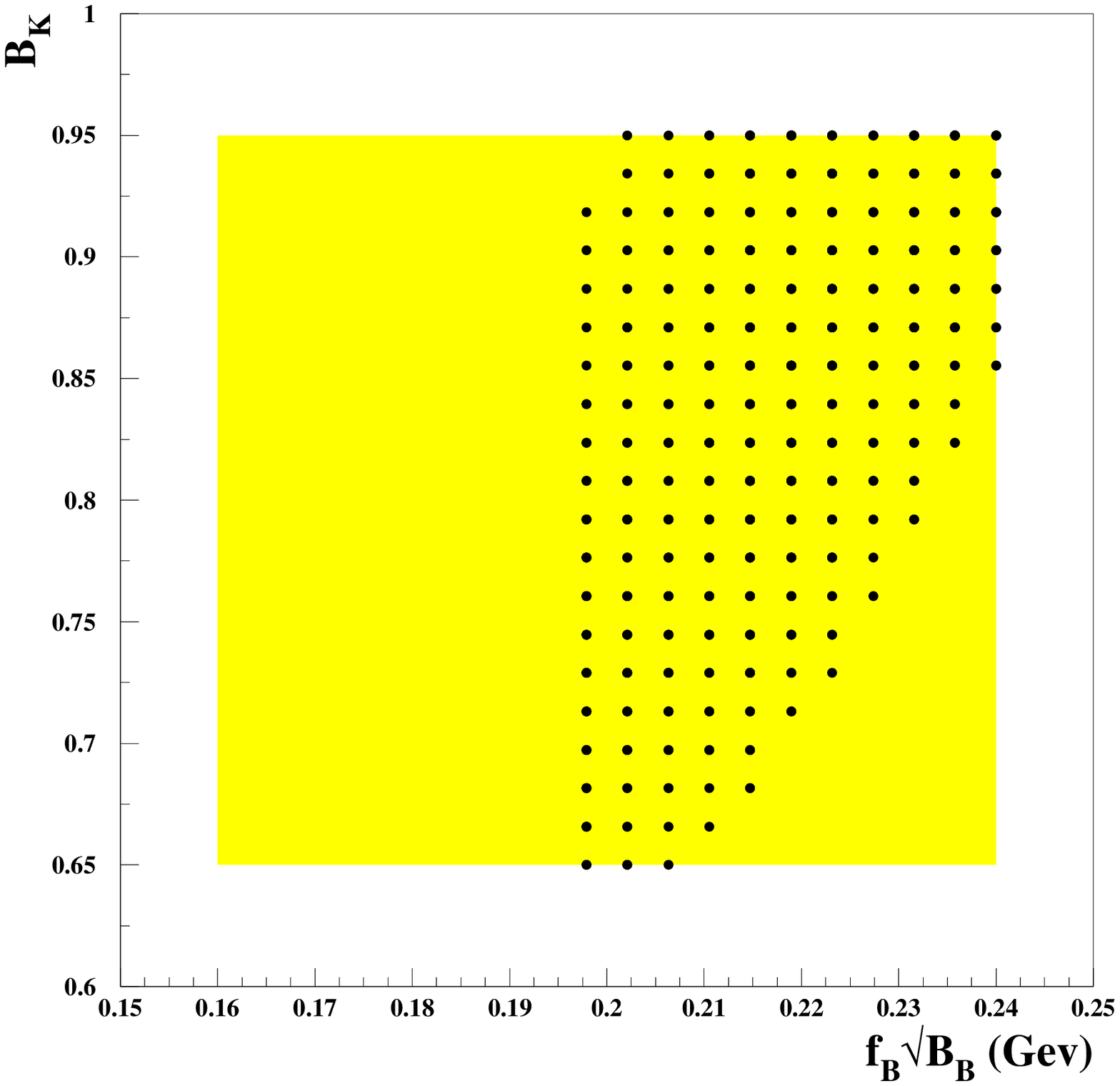,width=15cm}{\label{fig:fbb}The projection on the $\fbb,\bk$
plane of the theoretical models that survive the consistency cut of
5\% (points). The grey area depicts the whole range scanned.}

%%%%%%%%%%%%%%%%%%%%%%%%

\end{document}